\documentclass[a4paper]{article}

\usepackage{INTERSPEECH2021}

\usepackage{amsthm}
\usepackage{booktabs}
\usepackage{algorithm}
\usepackage{algorithmic}
\usepackage{multirow}

\title{Deep Annotation of Therapeutic Working Alliance in Psychotherapy}
\name{Baihan Lin$^1$, Guillermo Cecchi$^2$, Djallel Bouneffouf$^2$}
\address{
  $^1$Columbia University\\
  $^2$IBM Thomas J. Watson Research Center}
\email{baihan.lin@columbia.edu, gcecchi@us.ibm.com, djallel.bouneffouf@ibm.com}

\begin{document}

\maketitle
\begin{abstract}

The therapeutic working alliance is an important predictor of the outcome of the psychotherapy treatment. In practice, the working alliance is estimated from a set of scoring questionnaires in an inventory that both the patient and the therapists fill out. In this work, we propose an analytical framework of directly inferring the therapeutic working alliance from the natural language within the psychotherapy sessions in a turn-level resolution with deep embeddings such as the Doc2Vec and SentenceBERT models. The transcript of each psychotherapy session can be transcribed and generated in real-time from the session speech recordings, and these embedded dialogues are compared with the distributed representations of the statements in the working alliance inventory. We demonstrate, in a real-world dataset with over 950 sessions of psychotherapy treatments in anxiety, depression, schizophrenia and suicidal patients, the effectiveness of this method in mapping out trajectories of patient-therapist alignment and the interpretability that can offer insights in clinical psychiatry. We believe such a framework can be provide timely feedback to the therapist regarding the quality of the conversation in interview sessions. 

\end{abstract}

\section{Introduction}
A fundamental concept in psychotherapy is the working alliance between the therapist and the patient or, more generally, the client seeking help \cite{Bordin79}. The alliance involves several cognitive and emotional components of the relationship between these two agents, including the agreement on the goals to be achieved and the tasks to be carried out, and the bond, trust and respect to be established over the course of the therapy. Qualitative methods to quantify therapy outcomes led to the conclusion that the strength of the alliance is one of the main factors that predict success \cite{Wampold2015}.
Operational methods to quantify the alliance rely of evaluative reports by patients and therapists of whole sessions, typically limited to point-scales valuation \cite{horvath1981exploratory}. This approach does not make use of the nuances afforded by natural language, is time-consuming and difficult to follow through systematically outside of research studies; even more so is the evaluation of individual dialogue turns over the course of each session. Here we present an approach to quantify the degree of patient-therapist alliance by projecting each turn in a therapeutic session onto the representation of clinically established working alliance inventories, using language modeling to encode both turns
and inventories. This allows us not only to quantify the overall degree of alliance but also to identify granular patterns its dynamics over shorter and longer time scales. We also discuss how our approach may be used as a companion tool to provide feedback to the therapist and to augment learning opportunities for training therapists.

\section{Problem Setting}

\subsection{Working alliance analysis}

\begin{algorithm}[H]
% \small
 \caption{Working Alliance Analysis (WAA)}
 \label{alg:waa}
 \begin{algorithmic}[1]
 \STATE {\bfseries }\textbf{for} i = 1,2,$\cdots$, T \textbf{do}
\STATE {\bfseries } \quad Automatically transcribe dialogue turn pairs  $(S^p_i,S^t_i)$
\STATE {\bfseries }\quad \textbf{for} $(I^p_j, I^t_j) \in$ inventories $(I^p, I^t)$ \textbf{do} \\
\STATE {\bfseries }\quad \quad Score $W^{p_i}_{j}$ = similarity($Emb({I^p_j}), Emb(S^p_i)$) \\
\STATE {\bfseries }\quad \quad Score $W^{t_i}_{j}$ = similarity($Emb({I^t_j}), Emb(S^t_i)$) \\
\STATE {\bfseries } \quad \textbf{end for}
\STATE {\bfseries } \textbf{end for}
 \end{algorithmic}
\end{algorithm}

\begin{figure}[tb]
% \vspace{-0.8cm}
\centering
    \includegraphics[width=\linewidth]{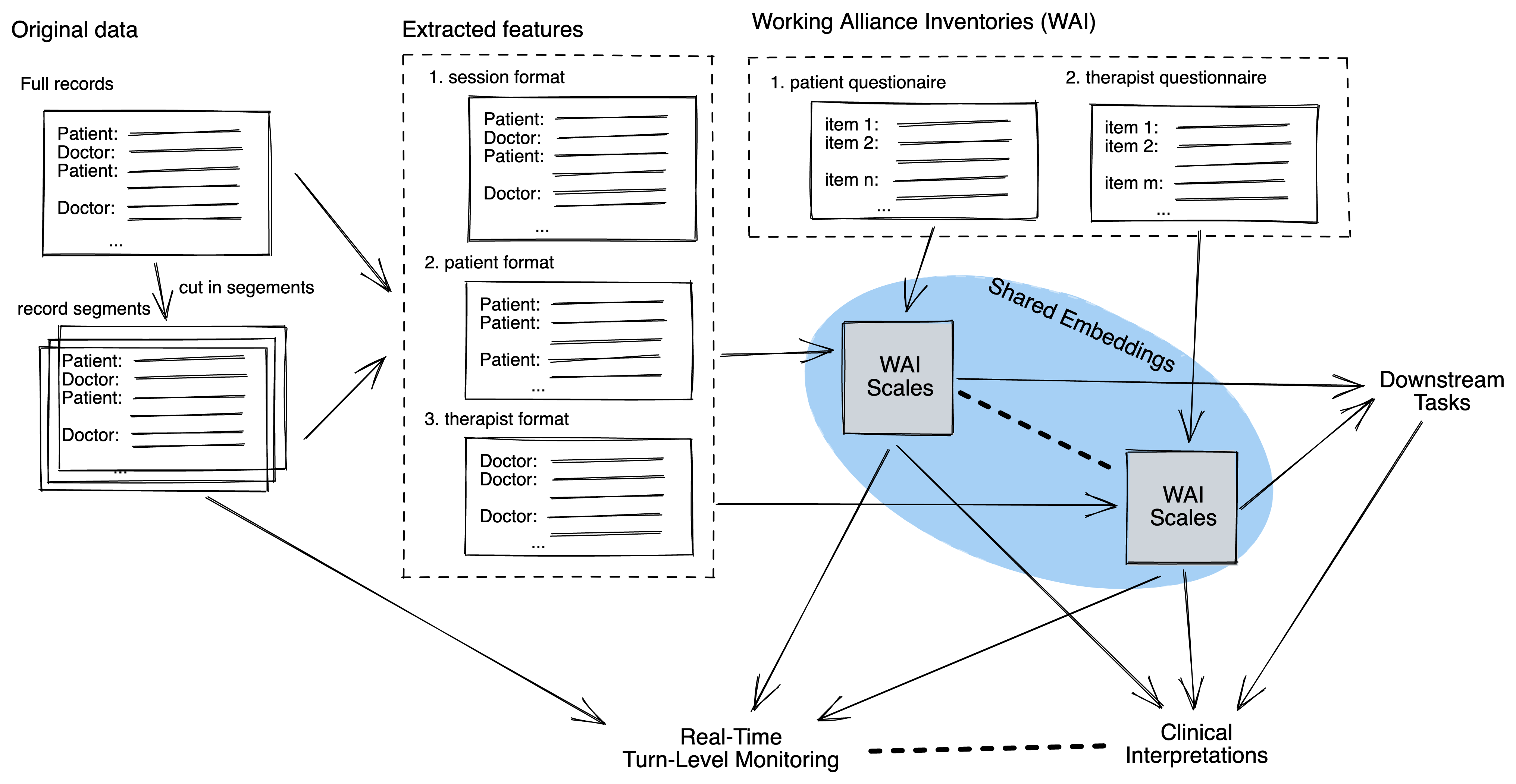}
\caption{Analytical Framework of working alliance analysis
}\label{fig:pipeline}
\end{figure}
 
The figure above is an outline of the analytic framework. We take the full records of a patient, or a cohort of patients belonging to the same condition. We either use it as is before the feature extraction, or we truncate them into segments based on timestamps or topic turns. As you can see, the original format is in pairs of dialogues. We can extract the features in three ways: first, we can use the full pairs of dialogues; second, we can only extract what the patient says; or the third option, we only extract what the doctor says. The three feature formats all have their pros and cons. The dialogue format contains all information, but the intents within the sentences come from two individuals, so they might mix together. The patient format contains the full narrative of the patients, which is usually more coherent, but it’s only part of the story. The therapist format, which people in computational psychiatry also believes to be some kind of semantic labels of what the patient feels, can be informative, but they can also be sometimes too simplistic.

\begin{figure}[tb]
% \vspace{-0.8cm}
\centering
    \includegraphics[width=\linewidth]{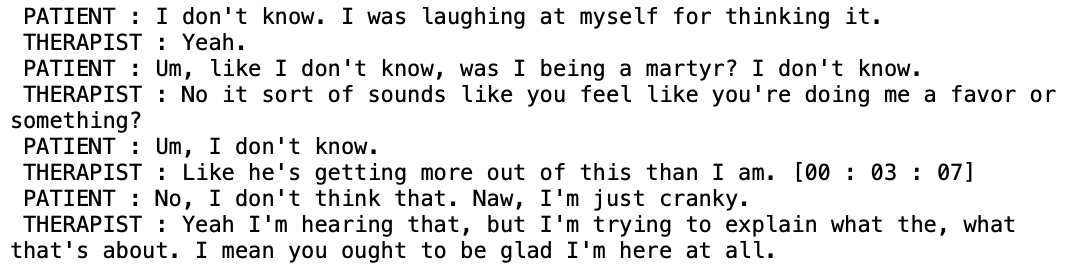}
\caption{Example dialogue from psychotherapy transcripts
}\label{fig:example}
\end{figure}

When we have the features, we compare the working alliance inventories with the embeddings. Algorithm \ref{alg:waa} outlines the process. During the session, the dialogue between the patient and therapist are transcribed into pairs of turns (such as the example in Figure \ref{fig:example}). We denote each patient response turn as $S^p_i$ followed by a therapist response turn $S^t_i$. They are treated as a dialogue pair. The inventories of working alliance questionnaires also come in pairs: $I^p$ for the patient (or client), and  $I^t$ for the therapist. They each consist of 36 statements. We embed both the dialogue turns and the inventories with deep sentence or paragraph embeddings, and then compute the cosine similarity between the embedding vectors of the turn and its corresponding inventory vectors. With that, for each turn (either by patient or by therapist), we obtain a 36-dimension working alliance score. We will describe in section \ref{sec:wai} the specific scales of our inferred working alliance scores which introduces interpretable information into our framework.

Here are a few downstream tasks and user scenarios that can plugged to our analytical frameworks. We can either use these extracted weighted topics to inform whether the therapy is going the right direction, whether the patient is going into certain bad mental state, or whether the therapist should adjust his or her treatment strategies. This can be built as an intelligent AI assistant to remind the therapist of such things.

\subsection{Psychotherapy Transcript Dataset}

The Alex Street \textit{Counseling and Psychotherapy Transcripts} dataset\footnote{https://alexanderstreet.com/products/counseling-and-psychotherapy-transcripts-series} consists of transcribed recordings of over 950 therapy sessions between multiple anonymized therapists and patients. This multi-part collection includes speech-translated transcripts of the recordings from real therapy sessions, 40,000 pages of client narratives, and 25,000 pages of reference works. These sessions belong to four types of psychiatric conditions: anxiety, depression, schizophrenia and suicidal. Each patient response turn $S^p_i$ followed by a therapist response turn $S^t_i$ is treated as a dialogue pair. In total, these materials include over 200,000 turns together for the patient and therapist and provide access to the broadest range of clients for our linguistic analysis of the therapeutic process of psychotherapy.

\section{Methods}

\subsection{Working Alliance Inventories}
\label{sec:wai}

\begin{figure}[tb]
% \vspace{-0.8cm}
\centering
    \includegraphics[width=\linewidth]{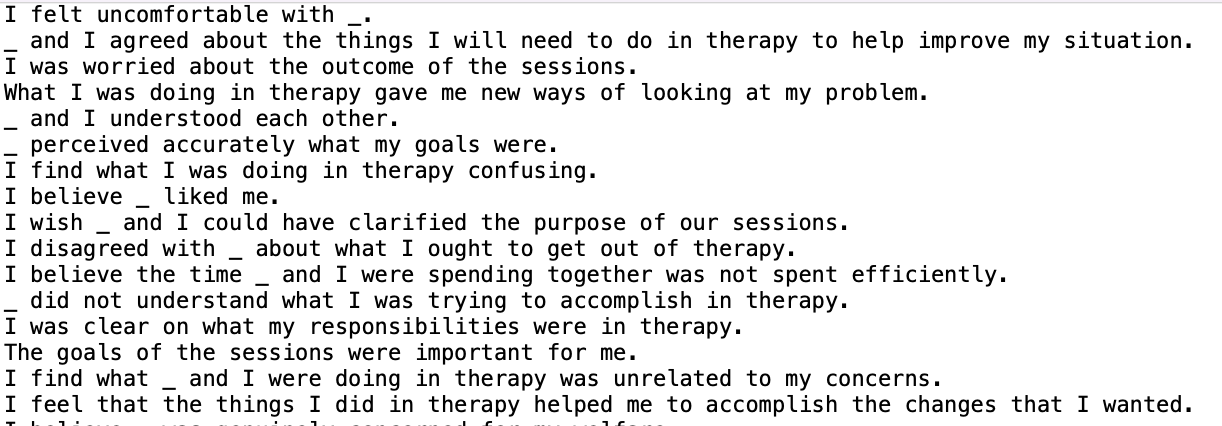}
\caption{Example statements in working alliance inventory
}\label{fig:wai}
\end{figure}

The Working Alliance Inventory (WAI) is a set of self-report measurement questionnaire that quantifies the therapeutic bond, task agreement, and goal agreement \cite{horvath1981exploratory,tracey1989factor,martin2000relation}. Since the original 12-item version \cite{tracey1989factor}, the inventory has used parallel versions for clients and therapist with good psychometric properties and helped establish the importance of therapeutic alliance in predicting treatment outcomes. The modern version of the inventory consists of 36 questions (Figure \ref{fig:wai}), and the participant is asked to rate each item on a 7-point scale (1=never, 7=always)\cite{martin2000relation}. The WAI aims to (1) measure alliance factors across all types of therapy, (2) document the relationship between the alliance measure and the corresponding theoretical constructs underlying the measure, and (3) related the alliance measure to a unified theory of therapeutic change \cite{horvath1994working}. 

\begin{figure}[tb]
% \vspace{-0.8cm}
\centering
    \includegraphics[width=\linewidth]{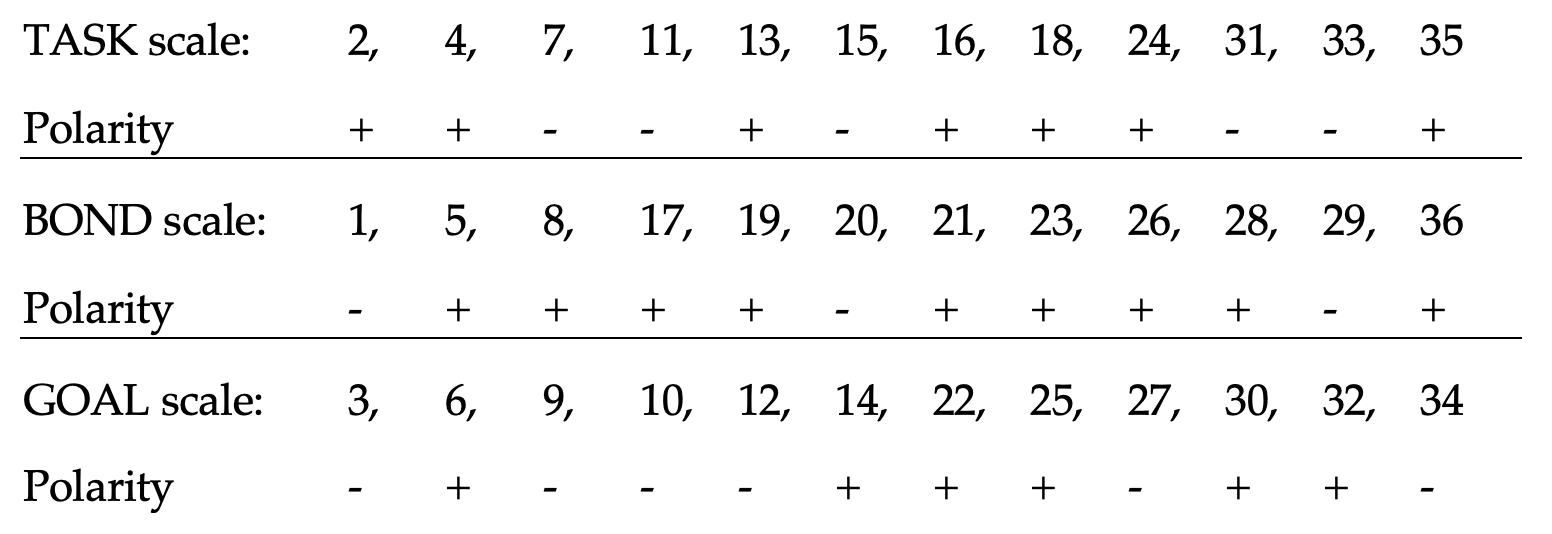}
\caption{Keys to the three scales of working alliance inventory
}\label{fig:wai_key}
\end{figure}

Operationally, the goal is to derive from these 36 items three alliance scales: the task scale, the bond scale and the goal scale. They measures the three major themes of psychotherapy outcomes: (1) the collaborative nature of the patient-therapist relationship; (2) the affective bond between therapist and patient, and (3) the therapist's and patient's capabilities to agree on treatment-related short-term tasks and long-term goals. The score corresponding to the three scales comes from a key table (Figure \ref{fig:wai_key}) which specifies the positivity or the sign weight to be applied on the questionnaire answer when summing in the end. The full scale is simply the sum of the scores of the three scales. The key table is like a weighting matrix that specifies the directionalities of the scales.

 \begin{figure}[tb]
% \vspace{-0.8cm}
\centering
    \includegraphics[width=\linewidth]{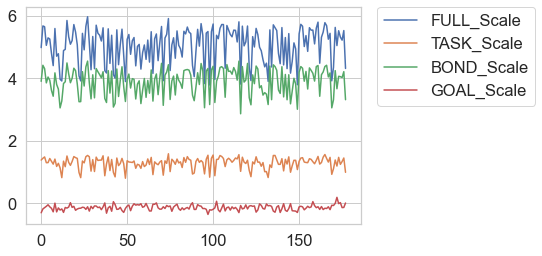}
\caption{Example trajectory of the working alliance scores
}\label{fig:example_traj_anxi100}
\end{figure}

 \begin{figure}[tb]
% \vspace{-0.8cm}
\centering
    \includegraphics[width=.49\linewidth]{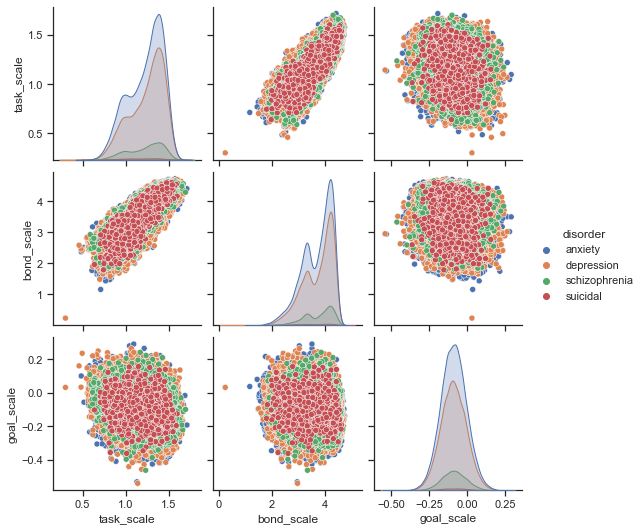}
    \hfill
        \includegraphics[width=.49\linewidth]{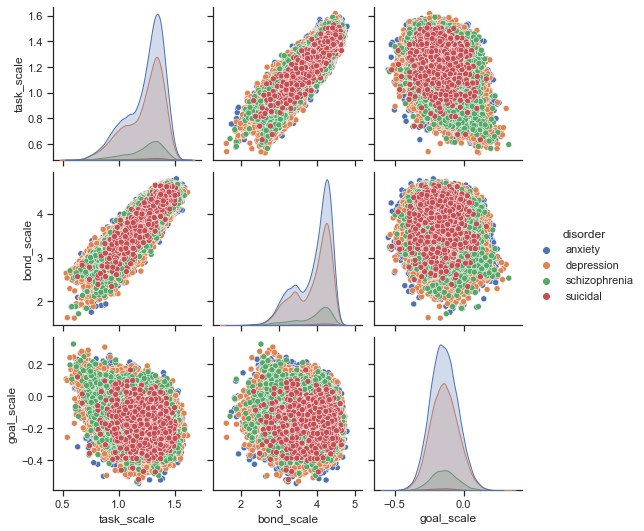}
\caption{Relational plots of the working alliance score scales (left: patient version; right: therapist version)
}\label{fig:rel}
\end{figure}

 \begin{figure}[tb]
% \vspace{-0.8cm}
\centering
    \includegraphics[width=\linewidth]{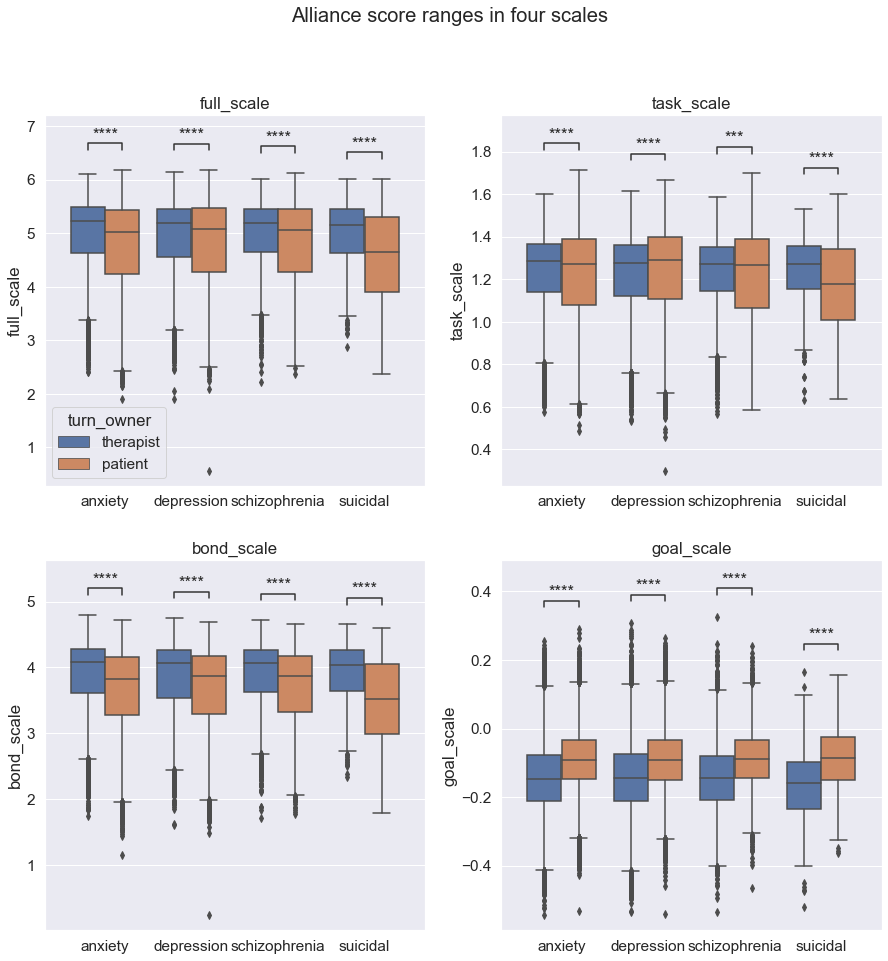}
\caption{Box plots of the working alliance scores
}\label{fig:box_scales}
\end{figure}

\subsection{Sentence Embeddings}

In principle, any sentence or paragraph embeddings can help us characterize the dialogue turns and inventories. In this work, we used two deep embeddings. The Doc2Vec embedding \cite{le2014distributed} is a popular unsupervised learning model that learns vector representations of sentences and text documents. It improves upon the traditional bag-of-words representation by utilizing a distributed memory that remembers what is missing from the current context. The other embedding we evaluated is the SentenceBERT \cite{reimers2019sentence}, which modifies a pretrained BERT network by using siamese and triplet network structures to infer semantically meaningful sentence embeddings. With these two deep embeddings, we embed the turn-level entries (either the dialogue turn in the transcripts, or the statement item in the working alliance inventories) into vectors of 300 or 384 dimensions. And then compute the cosine similarity between the vector at certain turn and an inventory entry.\footnote{Given the space limit, the results for the Doc2Vec are shown later, while the SentenceBERT results will be included in the extended online version as supplementary materials.}

\section{Results}

Figure \ref{fig:example_traj_anxi100} is an example time series of an anxiety psychotherapy session. We see that the alliance scores varies across the scales. If we investigate the relationship among the scales, we observe that the task scale positively correlates with the bond scale in both versions, while the goal scale slightly negatively correlates with the task scale in the therapist version (Figure \ref{fig:rel}).

We investigate the consistency of the alliance estimation by the patient vs. the therapist. Overall, comparing to the patient estimates, we observe that the therapist tends to overestimate the working alliance. More specifically, the therapists overestimates the task and bond scales, but underestimates the goal scale. These differences are all statistically significant ($p< 0.001$).

\begin{figure}[tb]
% \vspace{-0.8cm}
\centering
    \includegraphics[width=.55\linewidth]{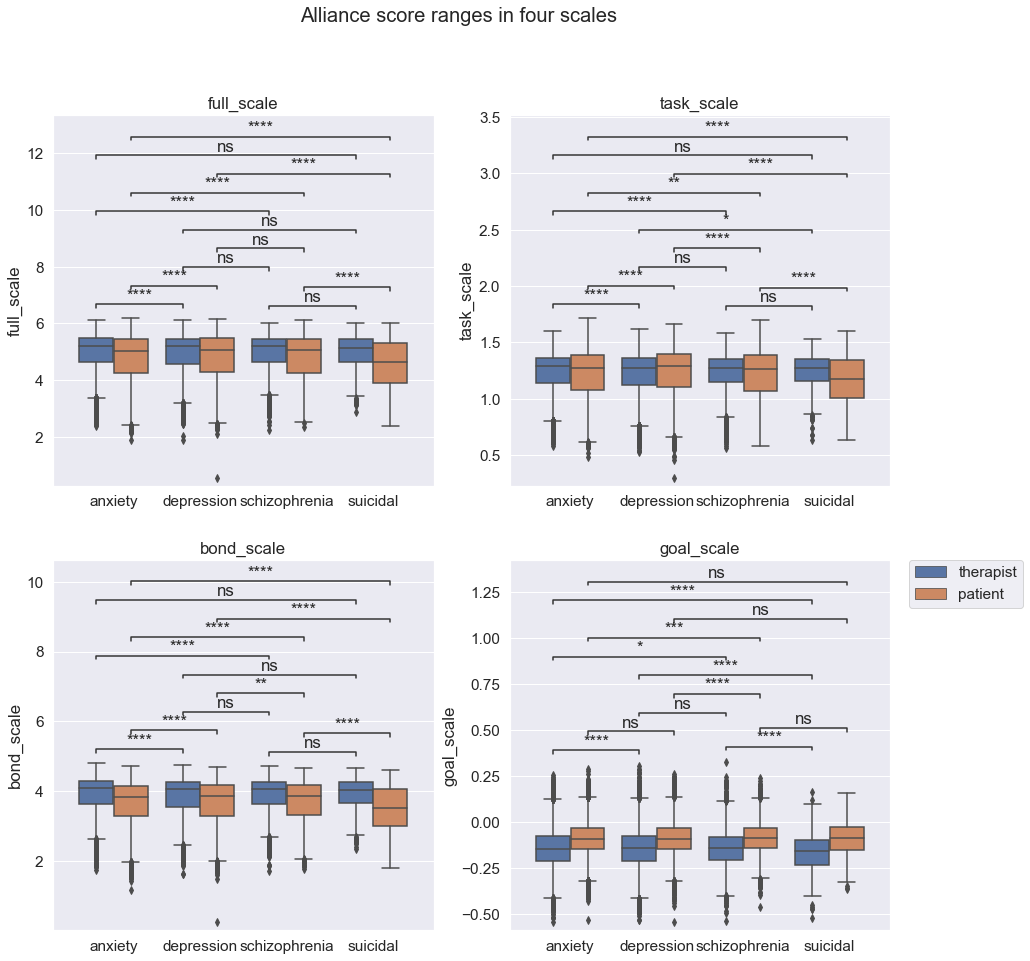}
    \includegraphics[width=.43\linewidth]{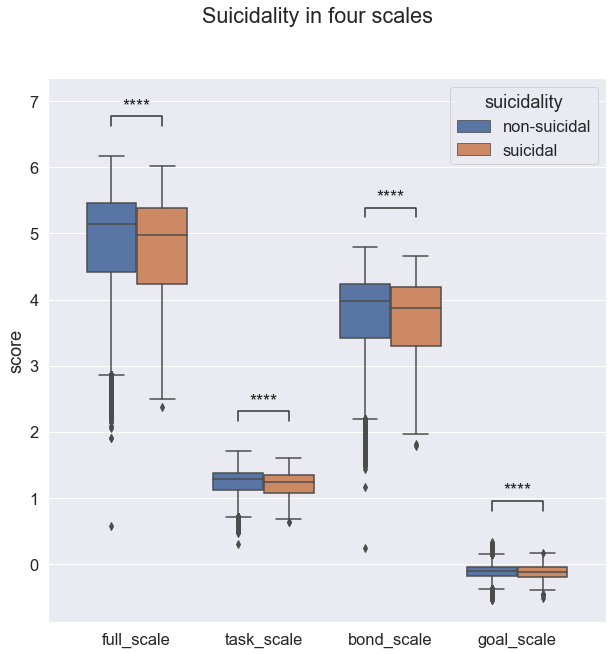}
\caption{Alliance scores across disorders
}\label{fig:box_disorders}
\end{figure}

Between the disorders, the alliance scores between anxiety and depression, and between anxiety and schizophrenia, are all significantly different in both the therapist and patient versions ($p< 0.001$). As in Figure \ref{fig:box_disorders}, the suicidality can be significantly detected based on the working alliance scores of all four scales. 

\begin{figure}[tb]
% \vspace{-0.8cm}
\centering
    \includegraphics[width=\linewidth]{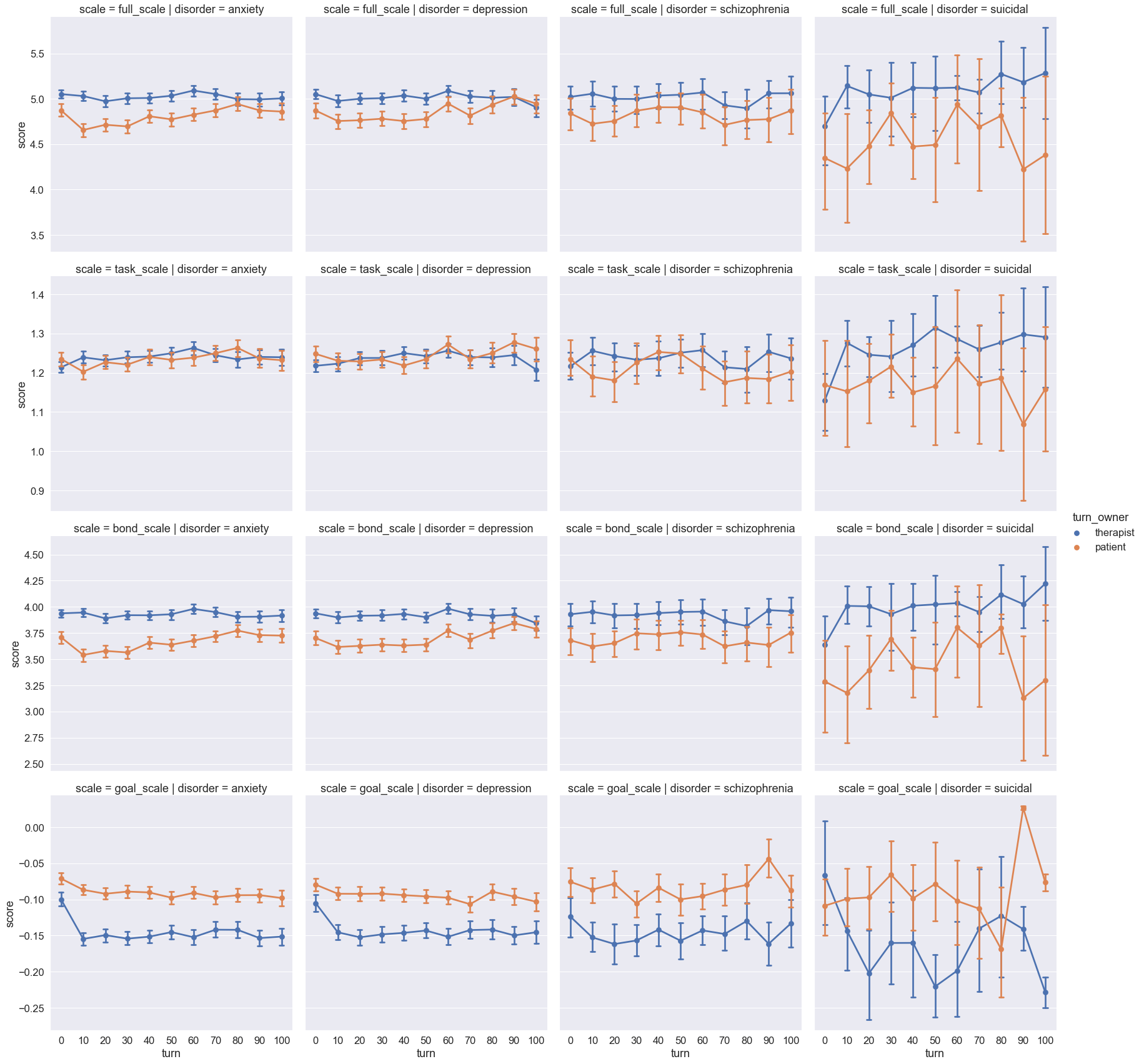}
\caption{Four-way ANOVA of the alliance dynamics
}\label{fig:anova_scales_disorders}
\end{figure}

 \begin{figure}[tb]
% \vspace{-0.8cm}
\centering
    \includegraphics[width=\linewidth]{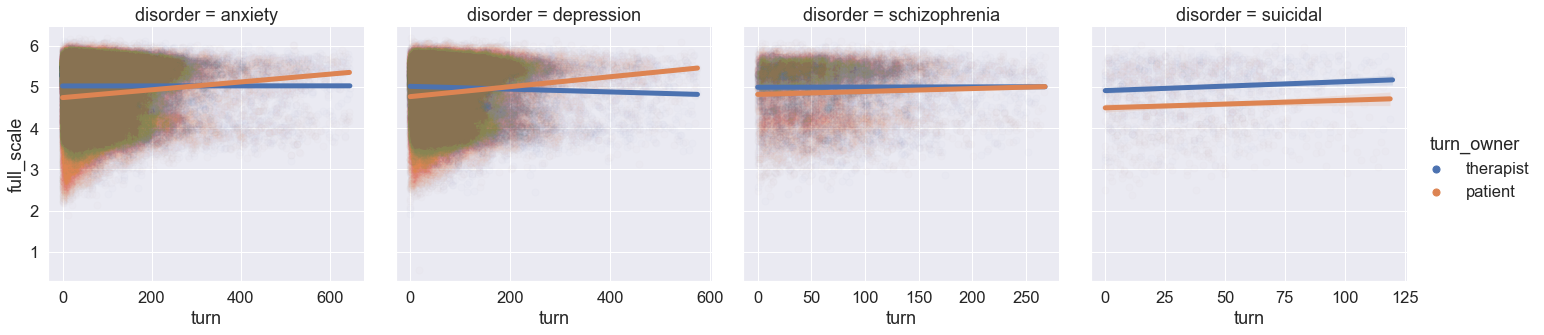}
\caption{Regression analysis of the temporal progression of the working alliance score
}\label{fig:reg_disorders}
\end{figure}

We also perform a four-way ANOVA upon the alliance scores as time-series sequences. Figure \ref{fig:anova_scales_disorders} demonstrates the difference of the dynamics of the therapeutic alliance across the psychiatric conditions. We observe that they vary by both the disorders and scales, and there appears to be certain trends along the temporal dimension (x-axis in each subplot). This is further supported by the linear regression analysis (Figure \ref{fig:reg_disorders}) that the patients with anxiety and depression have an upward alliance rating while their therapists tend to believe otherwise, and the therapists of the suicidal patients tend to have a higher alliance rating than their patients.  

 \begin{figure}[tb]
% \vspace{-0.8cm}
\centering
    \includegraphics[width=.95\linewidth]{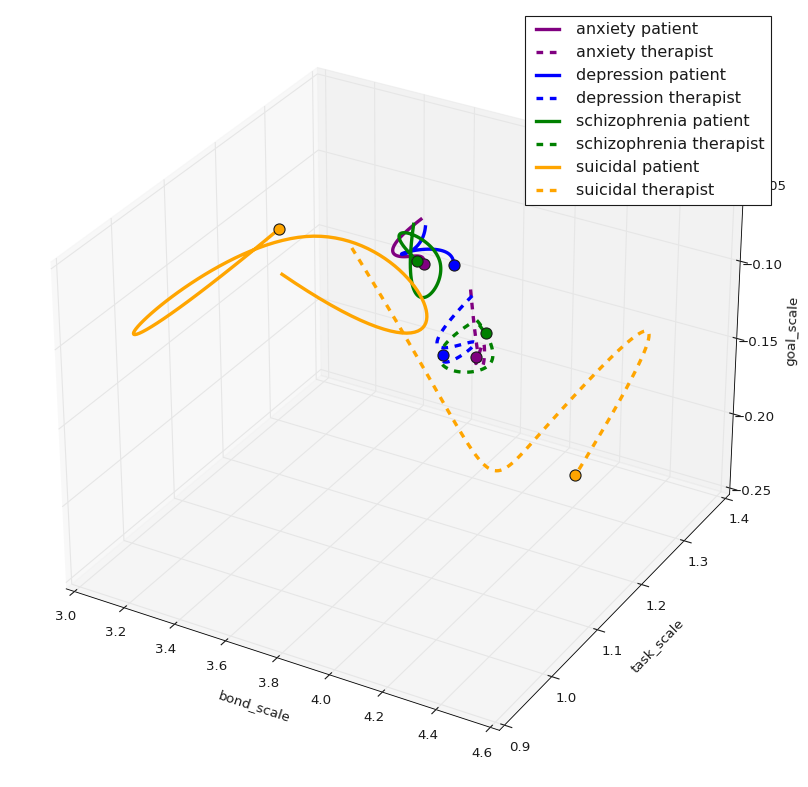}
\caption{The average 3d trajectories of different classes of psychiatric conditions in the alliance space (the dot meaning the end points of the trajectories) 
}\label{fig:trajs}
\end{figure}

We can also map out their trajectories in the alliance space of the three major scales (task, bond and goal). As in Figure \ref{fig:trajs}, we plot the average trajectories of different psychiatric conditions and notice that the suicidal trajectories are much more spread out in the bond and task scales (which aligns with the findings in the ANOVA plots). Based on the directionality, the suicidality trace shows a significant divergence trend. This is the first step of a potential turn-level resolution temporal analysis of the working alliance. We can be generalize in a sense that with this approach one can go over your sessions (as a therapist) and analyze the dynamics afterwards.

 \begin{figure}[tb]
% \vspace{-0.8cm}
\centering
    \includegraphics[width=\linewidth]{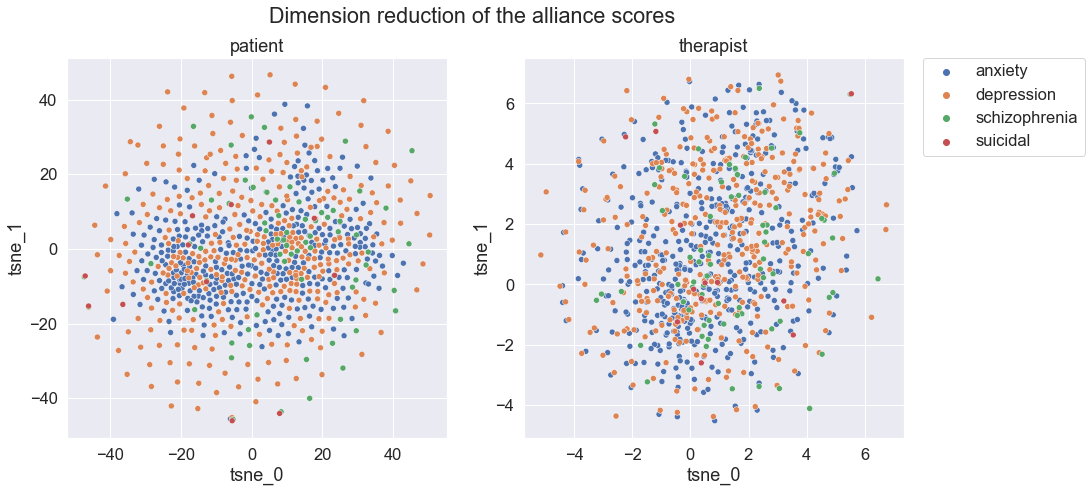}
\caption{Dimension reduction of the alliance trajectories
}\label{fig:dr}
\end{figure}

Given these time-series, we can visualize them with dimension reduction techniques such as t-SNE. Because the psychotherapy sessions come in different lengths, we compute the dynamic time wrapping distances between the session trajectories of 36-dimension alliance scores, and then use this pairwise distance matrix to perform the t-SNE unsupervised learning. Figure \ref{fig:dr} presents the difference between the manifolds of the therapist alliance trajectory space and the patient alliance space. We notice that the patient trajectories have two major clusters of alliance, while the therapist only has one. This is consistent with what we observed in the relational plots \ref{fig:rel} that the patient alliance scores in the task and bond scales follow a bi-modal distribution.

\begin{figure}[tb]
% \vspace{-0.8cm}
\centering
    \includegraphics[width=0.47\linewidth]{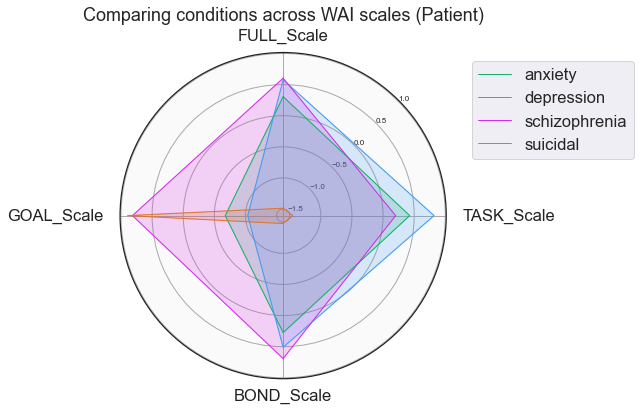}\hfill
    \includegraphics[width=0.47\linewidth]{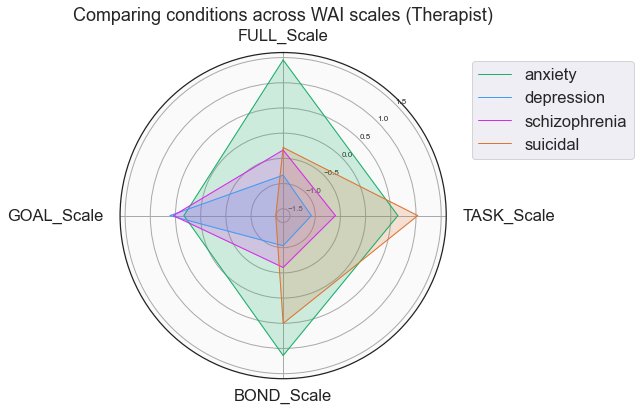}
\caption{Radar plots of the working alliance scores
}\label{fig:radar}
\end{figure}

We can further aggregate the alliance score by averaging all time points together all into the four scales. To plot the scale with respect to one another in a single plot, we normalize each scale to standard normal and present the radar plots of the scale-wise features of the patient and the therapist (Figure \ref{fig:radar}). We observe that the suicidal patient are comparatively most imbalanced, large only in the goal scale and small in all others. While, on the therapist version, it is the opposite, which aligns with the observation made in the 3d trajectories.

\section{Discussion}

Our analytic approach reveals several insightful features of the therapeutic relationship. We observe systematic differences in the mean inferred alliance scores between patients and therapists, and also across disorders. However the in-session evolution of the inferred scores provide a much more interesting perspective. In particular, while all conditions show a systematic misalignment of scores between patients and therapists, this is significantly starker for suicidality, something that can be observed in the mean as well as in the time trace for full and sub-scales. In contrast, anxiety and depression display a clear trend for the full and the bond scales to {\sl converge} as the sessions progress, something not present in the task and goal scales, nor in schizophrenia or suicidality. These features of the therapeutic dialogue can be mapped to what in psychiatry is usually called {\sl alignment} and plays an important symptomatic and diagnostic role in several neuropsychiatric conditions, e.g., in relation to the hypothesis of Theory of Mind for schizophrenia \cite{Janna2020}. By analyzing past sessions, and eventually sessions in real time, trained therapists may be able to identify key segments of the therapy leading to breakthroughs, compounding their expertise with further causal/predictive analytic modeling, while trainees may sharpen their intuition by reading or watching annotated versions of sessions conducted by experts. Needless to say, coupled with a generative language model and further statistical optimization, it may be possible to design limited chatbots to engage patients in triage and emergency response \cite{Garg2020}.

\section{Conclusions}

We have presented an approach that combines the state-of-the-art language modeling with the knowledge and practical expertise in psychotherapy, as captured in therapy-evaluation inventories, to provide a uniquely granular representation of the evolution of the interaction of patients and therapists. While here we focus specifically on the Working Alliance Inventory, our method is generic and can be extended to the broader spectrum of assessment instruments. Finally, it would be possible to refine and further validate the language-based estimation of working alliance by
providing punctuated rater evaluations as inference anchors. Next steps include predicting these inference anchors as states (like \cite{lin2022neural,lin2020predicting}) and training chatbots as reinforcement learning agents given these states (like \cite{lin2020story,lin2021models,lin2020unified}).

\clearpage
\bibliographystyle{IEEEtran}
\bibliography{main}

\end{document}